\documentclass{elsart}
\usepackage{graphicx,amssymb,amsmath,times}
\journal{Astroparticle Physics}
\begin{document}
\begin{frontmatter}
\title{Observations of TeV $\gamma$-rays from Mrk 421 during Dec. 2005 to Apr. 2006 with the TACTIC  telescope}
\author{K.K.Yadav\corauthref{cor}},
\corauth[cor]{Corresponding author.}
\ead{kkyadav@barc.gov.in}
\author{P.Chandra}, \author{A.K.Tickoo}, \author{R.C.Rannot}, \author{S.Godambe}, \author{M.K.Koul}, \author{V.K.Dhar}, \author{S.Thoudam}, \author{N.Bhatt}, \author{S.Bhattacharyya}, \author{K.Chanchalani}, \author{H.C.Goyal}, \author{R.K.Kaul},              \author{M.Kothari}, \author{S.Kotwal}, \author{R.Koul}, \author{S.Sahayanathan}, \author{M.Sharma}, \author{K.Venugopal}
\address {Astrophysical  Sciences  Division, Bhabha Atomic Research Centre. \\
Mumbai - 400 085, India.}
\begin{abstract}
The TACTIC $\gamma$-ray telescope has observed Mrk 421 on 66 clear nights from Dec. 07, 2005 to Apr. 30, 2006, totalling $\sim$ 202 hours of on-source observations. Here, we report the detection of flaring activity from the source  at $\geq$ 1 TeV energy and the time-averaged differential $\gamma$-ray spectrum in the energy range $1-11$ TeV for the data taken between Dec. 27, 2005 to Feb. 07, 2006 when the source was in a relatively higher state as compared to the rest of the observation period. Analysis of this  data spell, comprising about $\sim$97h reveals the presence of a $\sim 12.0 \sigma$ $\gamma$-ray signal  with  daily flux of $>$ 1 Crab unit  on several days. A pure power law spectrum with exponent $-3.11\pm0.11$ as well as a power law spectrum with an exponential cutoff $(\Gamma = -2.51\pm0.26$ and $E_0=(4.7\pm2.1) TeV)$ are found to provide  reasonable fits  to the inferred differential spectrum within statistical  uncertainties. We believe that the TeV light curve presented here, for nearly 5 months of extensive coverage, as well as the spectral information at $\gamma$-ray energies of $>$ 5 TeV provide a useful input for other groups working in the field of $\gamma$-ray astronomy.   
\end{abstract}

\end{frontmatter}
\section{Introduction}
Markarian 421 is the closest known TeV blazar (z=0.030) and was the first extragalactic source detected at TeV energies using imaging atmospheric Cerenkov telescopes [1,2]. The source has been regularly monitored by different groups since then [3-9]. It has been seen that the TeV $\gamma$-ray emission from Mrk 421 is highly variable with variations of more than one order of magnitude and ocassional flaring doubling time of as short as 15 mins [10,11]. Since its detection in the TeV energy range, Mrk 421 has also been  the target of several multiwavelength observation campaigns [12-15]. Several groups have also determined the energy spectrum  of Mrk 421, both at low average flux levels of $<$ 1 Crab Unit  and from intense flares of  $>$ 2 Crab Units. The recent results of these studies [4,5,7,16] suggest that  the spectrum is compatible with a power law combined with an exponential  cutoff.  It has also been reported  that the spectrum hardens as the flux increases [5,16], either because of an increase in the cutoff energy or a change in the spectral index itself.  Differences in the energy spectrum of Mrk 421 and Mrk 501  have also been addressed to understand the  $\gamma$-ray production mechanisms of these objects and  absorption effects at the source or in the intergalactic medium due to interaction of gamma-rays with the extragalactic background photons [17]. A recent review on  the observational  aspects of TeV $\gamma$-ray emission  from blazars can be found in [18].
\section{ Brief  description of the TACTIC  telescope}
The TACTIC (TeV Atmospheric Cerenkov Telescope with Imaging Camera) $\gamma$-ray telescope  has been set up at Mt. Abu ( 24.6$^\circ$ N, 72.7$^\circ$ E, 1300m asl), India for studying emission of  TeV  $\gamma$-rays  from celestial sources. The telescope deploys a F/1 type tracking light collector of $\sim$9.5 m$^2$ area, made up of 34 x 0.6 m diameter, front-coated spherical glass facets which have been prealigned to produce an on-axis spot of $\sim$ 0.3$^\circ$ diameter at the focal plane. The telescope uses a 349-pixel photomultiplier tube (ETL 9083UVB) -based imaging camera  with a uniform pixel resolution of $\sim$0.3$^\circ$ and a  field-of-view of $\sim$6$^\circ$x6$^\circ$  to record images of atmospheric Cerenkov events. The present data has been collected with inner 225 pixels (15$\times$ 15 matrix) of the full imaging camera. The innermost 121 pixels (11$\times$ 11 matrix) are used for  generating the event trigger, based on the NNP (Nearest Neighbour Pairs) topological logic [19], by demanding a signal $\geq$ 25  pe for the 2 pixels which participate in the trigger generation. Whenever the single channel rate of any two  or  more pixels in the trigger region goes outside the preset operational band (5 Hz-100 Hz), it is automatically restored to within the prescribed range by appropriately adjusting the high voltage of the pixels [20]. The resulting change in the photomultiplier (PMT) gain is monitored by repeatedly flashing a blue LED, placed at a distance of $\sim$1.5m from the camera. The advantages of using such a scheme are that in addition to  providing control over chance coincidence triggers, it also ensures  safe operation of PMTs with typical anode currents of $\leq$ 3 $\mu$A.  The back-end signal processing hardware of the telescope is based on medium channel density  NIM and CAMAC  modules developed inhouse. The data acquisition and control system of the telescope [21] has been designed around a network of PCs running the QNX (version 4.25) real-time operating system. The triggered events are digitized by CAMAC based 12-bit Charge to Digital Converters (CDC) which have a full scale range of 600 pC. 
The telescope has a pointing and  tracking accuracy of better than $\pm$3 arc-minutes. The tracking accuracy  is checked  on a regular basis  with so called "point runs", where an optical  star   having  its declination  close  to that of the  candidate $\gamma$-ray source is  tracked continuously  for about 5 hours.  The  point run calibration data  (corrected zenith and azimuth angle  of the telescope  when the star image is centered)  are  then incorporated in the telescope drive system software so that  appropriate corrections    can be  applied directly  in real time  while tracking  a candidate $\gamma$-ray source. 
\par
Operating at $\gamma$-ray  threshold energy of $\sim$1.2 TeV, the telescope records a  cosmic ray event rate of $\sim$2.0 Hz at a typical zenith angle of 15$^\circ$. The telescope has a 5$\sigma$ sensitivity of detecting Crab Nebula in 25 hours of observation time  and has so far   detected  $\gamma$-ray emission from the Crab Nebula, Mrk 421 and Mrk 501. Some of the results obtained on various candidate $\gamma$-ray sources are discussed in [22$-$24].   
\section{ Observations and data analysis}
Using  the TACTIC imaging telescope, Mrk 421  was observed  for $\sim$ 202 h  between Dec. 07,2005 to  Apr. 30, 2006. Only $\sim$30h of data was recorded in the off-source direction in order to maximize the on-source observation time and to improve the chances of recording possible flaring activity from the source. The total on-source data  has been  divided into 6 spells, where each spell  corresponds to one  lunation period. The zenith angle of the observations was $\leq$45$^\circ$. 
The general quality of the recorded data was checked by referring to the sky condition log  and  compatibility of the prompt and chance coincidence rates with Poissonian statistics. The imaging data recorded  by the telescope   was corrected for inter-pixel gain variation and  then subjected to the standard two-level image  'cleaning' procedure [25] with picture and boundary thresholds of  6.5$\sigma$ and 3.0$\sigma$, respectively. The  image  cleaning threshold levels were first optimized on the Crab data and then applied to the Mrk 421 data.  The clean Cherenkov images were characterized by calculating their standard image parameters like LENGTH, WIDTH, DISTANCE, ALPHA, SIZE and   FRAC2 [26,27]. The standard Dynamic Supercuts [28] procedure was then used to separate $\gamma$-ray  like images from the huge background of cosmic rays.   The $\gamma$-ray selection  criteria ( Table 1) used  in the  analysis  have been obtained  
\begin{table}[h]
\caption{ Dynamic Supercuts  selection  criteria used for analyzing the TACTIC data}
\centering
\begin{tabular}{|c|c|}
\hline 
Parameter  & Cut Values\\
\hline
LENGTH (L) & $0.11^\circ\leq L \leq(0.235+0.0265 \times \ln S)^\circ$\\
\hline
WIDTH  (W) & $0.06^\circ \leq W \leq (0.085+0.0120 \times \ln S)^\circ$\\
\hline
DISTANCE (D) & $0.52^\circ\leq D \leq 1.27^\circ cos^{0.88}\theta$ ;($\theta$$\equiv$zenith ang.)\\
\hline
SIZE (S)  & $S \geq 450 d.c$ ;(6.5 digital counts$\equiv$1.0 pe )\\
\hline
ALPHA ($\alpha$) &  $\alpha \leq 18^\circ$\\
\hline
FRAC2 (F2) &  $F2 \geq 0.35$ \\
\hline
\end{tabular}
\end{table}  
on the basis of  dedicated  Monte Carlo simulations carried out for  the TACTIC telescope.   The cuts have   also been  validated further  by applying  them  to the Crab Nebula data  collected by the telescope.
\section{ Results -- Alpha plot analysis and light curve}
A well established  procedure to extract the $\gamma$-ray  signal from the cosmic ray background using single imaging telescope is to plot the  frequency distribution of  ALPHA parameter (defined as the  the angle between the major axis of the image and the line between the image centroid and camera center) of shape and DISTANCE selected events. This  distribution is expected to be flat  for the isotropic background of cosmic events. For $\gamma$-rays, coming from a point source, the distribution is expected to show a peak  at  smaller $\alpha$ values. Defining $\alpha\leq18^\circ$ as the $\gamma$-ray domain and $27^\circ\leq\alpha\leq81^\circ$ as the background region, the number of $\gamma$-ray events is then calculated  by subtracting the expected number of background events (calculated  on the basis of background  region) from the  $\gamma$-ray domain events. Estimating the expected background level in the $\gamma$-domain  by following this approach  is  well known  [29] and has been  used quite extensively by  other groups  when  equal amount of off-source data is not available. However, we  have also  validated  this  method  for the TACTIC telescope by using separate off-source data  on a regular basis and the $\alpha$ distribution of this data in range $\alpha\leq81^\circ$ is in good agreement with the expected flat distribution. The significance of the excess events has  been finally calculated by using the maximum likelihood ratio method of Li $\&$ Ma [30].
\par
In order to test the validity of the data analysis chain and, in particular, the energy estimation procedure, we have first analyzed the Crab Nebula data collected  by the TACTIC imaging telescope  for $\sim$101.44h during Nov. 10, 2005 - Jan. 30, 2006. The events selected after using the Dynamic Supercuts procedure (Table 1) yield an excess of $\sim$(839$\pm$89) $\gamma$-ray events with a statistical significance of $\sim$9.64$\sigma$. The corresponding average $\gamma$-ray rate turns out to be $\sim$(8.27$\pm$0.88)h$^{-1}$. The same data sample has been analyzed again after restricting the zenith angle of the observations to  15 $^\circ$ - 45$^\circ$  (similar to the zenith angle range which  Mrk 421 would cover) so that the resulting $\gamma$-ray rate  can be designated as  a reference of 1 Crab Unit (CU)  while  interpreting  the Mrk 421  data. The analysis yields an excess of $\sim$ (598$\pm$69) $\gamma$-ray events in an observation time of $\sim$63.33h with a corresponding $\gamma$-ray rate of $\sim$(9.44$\pm$1.09)h$^{-1}$, leading to  the conversion: 1 CU$\equiv$(9.44$\pm$ 1.09)h$^{-1}$. While one would have expected this rate to decrease because of increase in the threshold energy of the telescope at higher zenith angles, the  reason behind this increase  is the  superior $\gamma$-ray acceptance of Dynamic Supercuts at higher zenith angles which overcompensates the threshold change effect.  
\par
In the case of  Mrk 421, we have first analyzed the data for each spell separately.  Furthermore, it has also been ensured that the data on both sources ( Crab and Mrk 421) are subjected to exactly similar analysis procedures to avoid  any  source dependent bias while determining  the energy spectrum of Mrk 421. The results  of the  spell wise  analysis are presented in Table 2. An examination of this table clearly indicates that Mrk 421  was most active  during  spell 2 and spell 3 observations.  
\begin{table}[h]
\caption{ Detailed  Spell wise  analysis report of  Mrk 421 data}
\centering
\begin{tabular}{|c|c|c|c|c|c|c|}
\hline 
Spell & Obs. time  &$\gamma$-ray  & $\gamma$-ray   &  Signifi-         & $\chi^2$ /dof  &   Prob  \\ 
      &  ( hrs. )  & events      & rate (h$^{-1}$) &  cance($\sigma$)  & (27$^\circ$ $\leq$$\alpha$$\leq$81$^\circ$)               &         \\
\hline
 I    &    9.24   & 9 $\pm$25    & 0.97 $\pm$2.67    & 0.37              & 2.97 /5        & 0.705  \\
\hline 
 II   &   35.71   & 275$\pm$49   & 7.70 $\pm$1.37    & 5.79              & 5.77 /5        & 0.329  \\
\hline
 III   &   61.53   &676 $\pm$66  & 10.99 $\pm$1.07  & 10.64              & 8.57 /5        & 0.127  \\
\hline
 IV   &    34.54   & 91 $\pm$47  & 2.64$\pm$1.37     & 1.94              & 5.10 /5        & 0.404  \\ 
\hline
 V     &   31.14   & 61$\pm$38   & 1.96 $\pm$1.23    & 1.61              & 2.44 /5        & 0.785  \\
\hline
 VI    &   29.55   & 123$\pm$33  & 4.16 $\pm$1.11   & 3.86              & 3.01 /5        &  0.698  \\
\hline
\hline
 All data &  201.72 & 1236$\pm$110  & 6.13 $\pm$ 0.55  & 11.49           & 4.57 /5        & 0.471  \\
\hline
 II +III  &  97.24  & 951$\pm$82   & 9.78 $\pm$0.84   & 12.00            & 5.00 /5        &  0.416  \\
\hline
\end{tabular}
\end{table}   
Fig. 1a  gives the $\alpha$-distribution when all the data collected for $\sim$  201.72 h ( Dec. 07, 2005 - Apr. 30, 2006, spell 1 to spell 6) is analysed together.  
\begin{figure}[h]
\centering
\includegraphics*[width=0.95\textwidth,angle=0,clip]{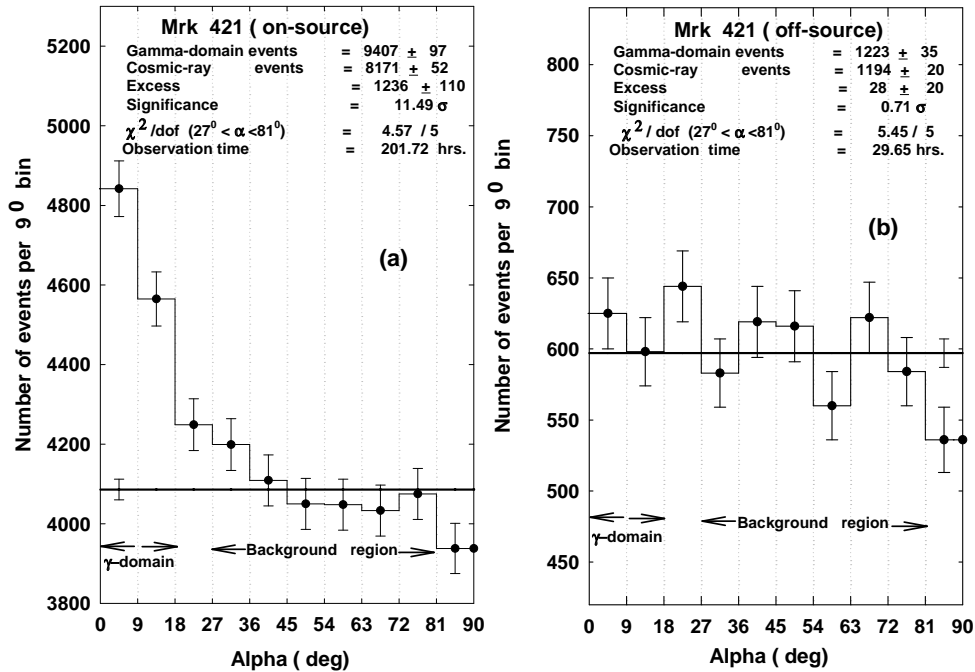}
\caption { (a) On-source Alpha  plot  for Mrk 421  when all the data  collected  between Dec. 07, 2005 - Apr. 30, 2006  for  $\sim$ 201.72 h  is analysed. (b) Off-source Alpha  plot for $\sim$ 29.65 hrs.  of  observation time. The horizontal lines in these figures indicate the expected background in the $\gamma$-domain obtained by using the background region ($27^\circ\leq\alpha\leq81^\circ$).}
\end{figure} 
The total data  yields an excess of $\sim$(1236$\pm$110) $\gamma$-ray  events with a statistical significance of $\sim$11.49$\sigma$. Fig. 1b  gives the corresponding $\alpha$-distribution for $\sim$ 29.65 h  of off-source data.  The procedure for estimating the expected background in the $\gamma$-domain by using the background region is also consistent with the results from the off-source  alpha plot (Fig.1b), when  its $\gamma$-ray domain events ($\alpha\leq18^\circ$ events) are appropriately  scaled up (to account for the difference in the on-source and off-source observation time) and compared with  the $\gamma$-ray domain events of Fig.1a. The results of this  calculation  yield a value of $\sim$(8320$\pm$238) events for the background level which is in close agreement with the value of $\sim$(8171$\pm$52) events obtained when background region of Fig.1a is used itself. It is worth mentioning  that $\chi^2$/dof of the background region (indicated by column 5 of Table 2) is also consistent with the assumption that the background  region is flat and thus can be reliably used for estimating the background level in the  $\gamma$-domain.
\par
The full light curve of Mrk 421 as recorded by the TACTIC imaging telescope is shown in Fig.2a, where  daily rates have been plotted as a function of modified  Julian date. 
\begin{figure}[h]
\centering
\includegraphics*[width=0.95\textwidth,angle=0,clip]{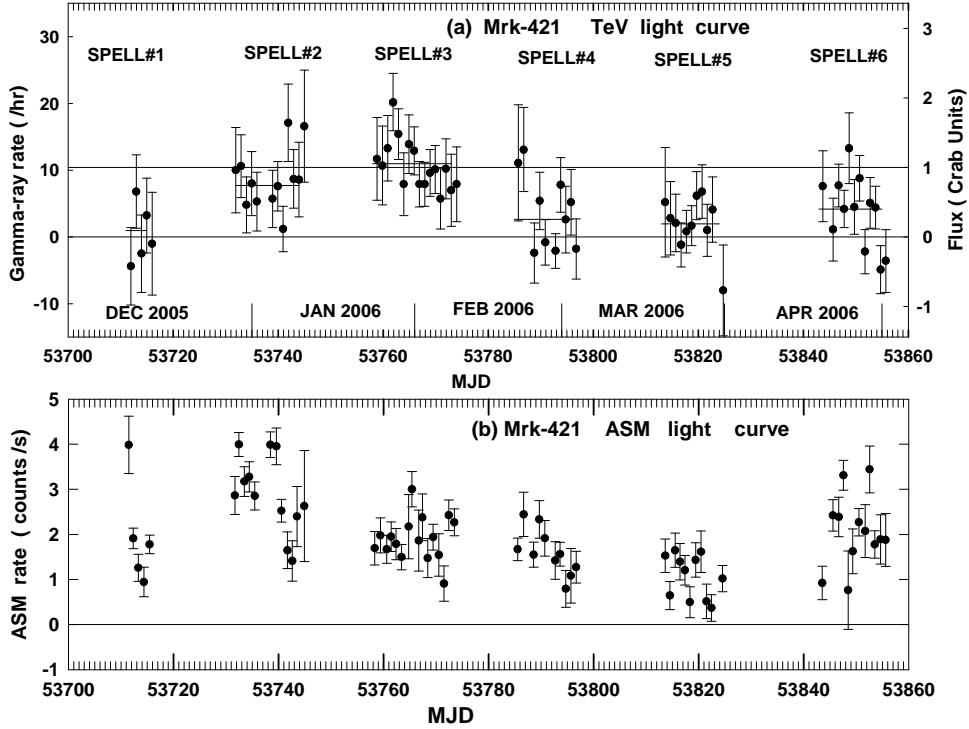}
\caption {(a) Light curve for Mrk 421 as recorded by the TACTIC imaging telescope Dec. 07, 2005 - Apr. 30, 2006 for a total observation time of $\sim$201.72h. The horizontal lines marked in between various flux points indicate the spell wise average $\gamma$-ray rate. (b) Daily plot of the average count rate in the 2-10 keV energy range as determined by RXTE/ASM  X-ray detector}
\end{figure} 
The  corresponding  light curve  showing  the daily  average count rate in the 2-10 keV energy band as determined by RXTE/ASM  [31]  is shown in Fig. 2b.  Since the ASM observations were not  simultaneous with the observations taken with the TACTIC telescope,  a detailed  correlation analysis  is  rather  difficult. We have also examined day-to-day flux variability in the TeV light  curve of the source for spell 2 and  spell 3 data. The  results   indicate  that we cannot claim any statistically  significant flux  variation  on a nightly basis because of rather large error bars.  
\section{ Monte Carlo simulations for energy reconstruction of $\gamma$-rays}
The Monte Carlo  simulation data  used  for developing  a  procedure  for $\gamma$-ray event selection and energy reconstruction of $\gamma$-rays  are  based on the CORSIKA (version 5.6211) air-shower  simulation code [32]. The simulated data-base for $\gamma$-ray showers uses about 34000 showers in the energy range 0.2-20 TeV  with an impact parameter of 5-250m. These showers have been generated at 5 different zenith angles ($\theta$= 5$^\circ$, 15$^\circ$, 25$^\circ$, 35$^\circ$ and  45$^\circ$). Furthermore, a data-base of  about 39000 proton initiated  showers in the energy range 0.4-40 TeV, distributed isotropically within a field-of-view 6$^\circ$x6$^\circ$, have also been  generated for studying the  gamma/hadron separation capability of Dynamic Supercuts and confirming the matching of experimental and simulated image parameter distributions. A supplementary code has been developed for the ray tracing of the Cerenkov photons [33] and also to take into account the wavelength dependent atmospheric absorption, the spectral response of the PMTs and the reflection coefficient of mirror facets and light cones. The Cerenkov  photon data-base, consisting  of number of photoelectrons registered by each pixel,  has been subjected to noise injection, trigger condition check and image cleaning. The resulting two dimensional 'clean' Cerenkov image of each triggered event is then used to determine various image parameters. The same simulation data base has also been used, as per the well known standard procedure [28], for calculating the effective area of $\gamma$-rays as a function of energy and zenith angle and,  also the $\gamma$-ray retention factors when Dynamic Supercuts  are applied to the simulated data.  Here, we  present only the details of our  energy reconstruction procedure  which  we believe  has been  attempted  for the first time.  
\par
Given the inherent power of ANN (Artificial Neural Network) to effectively handle the multivariate data fitting, we have developed an ANN-based energy estimation procedure  for determining the energy spectrum of a candidate $\gamma$-ray source.  An   ANN  is a non-linear statistical  data modeling  tool  which can be used to model complex relationships between inputs and outputs. The tool can be applied to problems   like  function approximation or regression analysis (similar to  what is being attempted here), classification (pattern recognition) etc. Although not for energy estimation, the idea of applying ANN to imaging telescope  data was attempted for the first time by Reynolds  and Fegan  [34].  While  a  detailed description  of  ANN  methodology  can be  found  in [34,35],  we will  only present  here the main steps  taken by us  to ensure reliability of end results. 
\par
The $\gamma$-ray energy reconstruction with a single imaging telescope,  in general, is a function of image SIZE, DISTANCE and zenith angle.The procedure followed  by us uses a 3:30:1(i.e 3 nodes in the input layer, 30 nodes in hidden layer and 1 node in the output layer) configuration of the ANN with resilient back propagation  training algorithm  [36] to estimate the energy of a $\gamma$-ray like event on the basis of its image SIZE, DISTANCE and zenith angle. The training and testing of the ANN  was done  in accordance with  the standard procedure  of dividing the  data base into two parts so that  one part could be used for the training and the remaining for testing.  We used about 10,000  events out of a total of 34000 parameterized  $\gamma$-ray images for training. The  3 nodes in the input layer correspond to   zenith angle, SIZE and DISTANCE, while the 1  node  in the output layer  represents  the expected  energy  (in TeV) of the event. In order to make  ANN training easier and to smoothen inherent event to event fluctuations,  we first  calculated the $<$SIZE$>$ and $<$DISTANCE$>$ of the training data sample of 10000 images by clubbing  together showers of a particular energy in various core distance bins  with  each bin having a size of 40m. Once satisfactory training of the ANN was achieved, the corresponding ANN generated weight-file was then used as a part of the main data analysis program so that the energy of a $\gamma$-ray like  event could be predicted without  using the ANN software package. Rigorous checks  were  also performed to ensure that Network did not become "over-trained" and the configuration used was properly optimized  with respect to number of iterations and number of nodes in the hidden layer.  Optimization of the network was done  by  monitoring the RMS error while training  the  ANN. The  optimized configuration  (3:20:1   with  5000 iterations) yielded a final  RMS error of $\sim$ 0.027 which  reduced only marginally  when the number of nodes in the hidden layer  or the number of iterations were  increased further.
\par
A plot of energy reconstruction  error obtained  for test data sample of 24000 parameterized images  is shown in Fig.3a. This plot has been obtained  by testing the ANN  with individual parameterized images at different energies within 5$^\circ$-45$^\circ$ zenith angle range.   Denoting the estimated energy by E$_{ANN}$ and the actual energy used during  Monte Carlo simulations by E$_{MC}$, Fig. 3b shows the frequency distribution of (ln(E$_{ANN}$/ E$_{MC}$) for all events shown in  Fig.3a  alongwith a Gaussian fit. 
\begin{figure}[h]
\centering
\includegraphics*[width=0.95\textwidth,angle=0,clip]{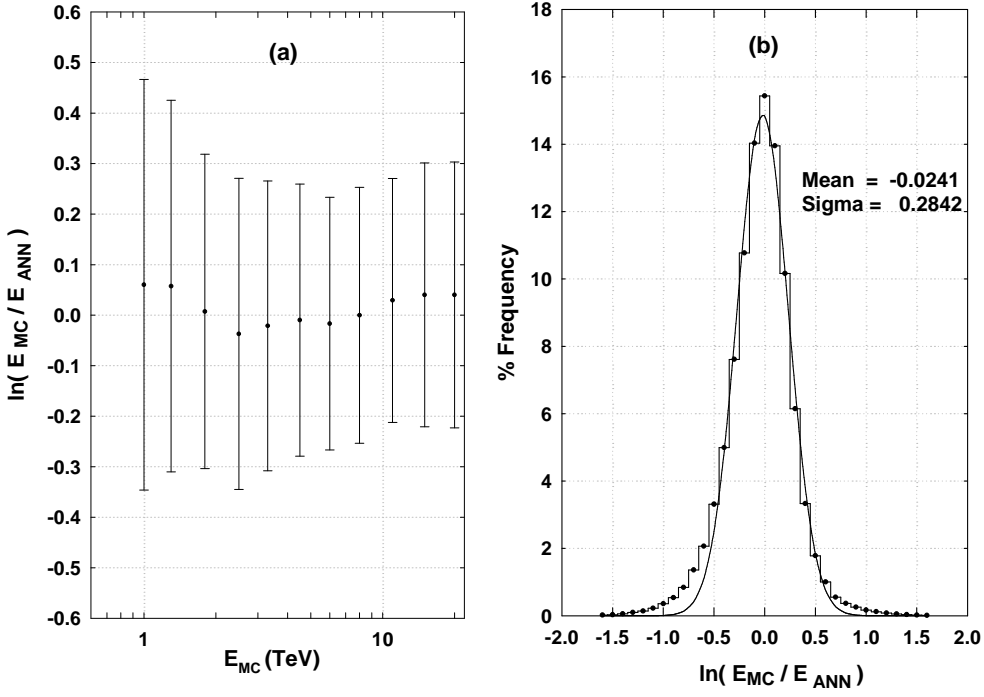}
\caption { (a) Error in the energy resolution function ( ln (E $_{ANN}$/ E$_{MC}$)) as a function of the  energy of the $\gamma$- rays for the simulated  showers generated  within  5$^\circ$-45$^\circ$  zenith angle range. (b) Frequency distribution of ( ln (E $_{ANN}$/ E$_{MC}$) for all events shown in Fig.1a  along with a Gaussian fit. }
\end{figure} 
Having a  value of $\sigma$(ln E) $\sim$ 28.4$\%$  (Fig.3b),  directly implies that  the  procedure should allow  us to retain  $\sim$84$\%$  $\gamma$-rays in most of the energy bins from $1-20$ TeV, if 6 bins per energy decade (i.e $\sigma$(ln E)) of $\sim$40$\%$ ) are used for determining the energy spectrum of a candidate $\gamma$-ray source.    
The  performance of the ANN-based energy reconstruction  procedure was also compared with the results obtained  
from the linear  least square fitting method. This method yielded a $\sigma$(ln E) of $\sim$35.4$\%$.  
It is also worth  mentioning here that the new ANN-based energy reconstruction method used here, apart from yielding a lower  $\sigma$(ln E) of $\sim$ 28.4$\%$ as compared to  $\sigma$(ln E) of $\sim$ 36$\%$ reported by the Whipple group [28], has the added advantage that it considers zenith angle dependence  of SIZE and DISTANCE  parameters as well. The procedure thus allows data collection over  a much wider zenith angle range as against a  coverage of upto 35$^\circ$ in case the zenith angle dependence is to be ignored. \\
\section{ Energy  spectrum of Mrk 421}
The differential photon flux  per energy bin  has been computed using the formula
\begin{equation}
\frac{d\Phi}{dE}(E_i)=\frac {\Delta N_i}{\Delta E_i \sum \limits_{j=1}^5 A_{i,j} \eta_{i,j} T_j}
\end{equation}
where $\Delta N_i$ and $d\Phi(E_i)/dE$ are the number of events and the differential flux at energy $E_i$, measured in the ith  energy bin $\Delta E_i$ and over the zenith angle range of 0$^\circ$-45$^\circ$, respectively. $T_j$ is the observation time in the jth zenith angle bin with corresponding energy-dependent effective area ($A_{i,j}$) and $\gamma$-ray acceptance ($\eta_{i,j}$). The 5 zenith angle bins (j=1-5) used are 0$^\circ$-10$^\circ$, 10$^\circ$-20$^\circ$, 20$^\circ$-30$^\circ$, 30$^\circ$-40$^\circ$  and 40$^\circ$-50$^\circ$ with  simulation data  available at 5$^\circ$, 15$^\circ$, 25$^\circ$, 35$^\circ$ and 45$^\circ$. The number of $\gamma$-ray events  ($\Delta N_i$)  in a particular  energy bin is  calculated  by subtracting the expected number of background events, from the  $\gamma$-ray domain events.
\par
In order to test the validity of the energy estimation procedure, we first used Crab Nebula  data collected by  the TACTIC  imaging telescope  for $\sim$101.44 h between Nov. 10, 2005 - Jan. 30, 2006. The $\gamma$-ray differential  spectrum  obtained   after applying the Dynamic Supercuts  and  appropriate values of  effective collection area and $\gamma$-ray acceptance  efficiency  (along with  their  energy and zenith angle dependence) is shown in Fig.4a. While  determining  the energy spectrum, we used the excess  noise factor method for converting the  image size in CDC counts to number of photoelectrons. The analysis of relative calibration data yields  a value of  1pe $\cong$  (6.5$\pm$1.2) CDC for this when an average value of $\sim$1.7 is used for excess noise factor of the photomultiplier tubes.  The  differential energy spectrum of the Crab Nebula  shown in Fig.4a  is  a  power law fit  $(d\Phi/dE=f_0 E^{-\Gamma})$  with  $f_0=(2.74\pm0.19)\times 10^{-11} cm^{-2}s^{-1}TeV^{-1}$  and $\Gamma=2.65\pm0.06$. The fit  has a $\chi^2/dof=0.53/6$ with a corresponding probability of 0.997. The errors in the flux constant and the spectral index are standard errors. Excellent matching of this spectrum with that obtained  by the Whipple and  HEGRA  groups [37,38]  reassures  that the procedure followed by us for obtaining the energy spectrum  of a $\gamma$-ray  source is quite  reliable.
\begin{figure}
\centering
\includegraphics*[width=0.95\textwidth,angle=0,clip]{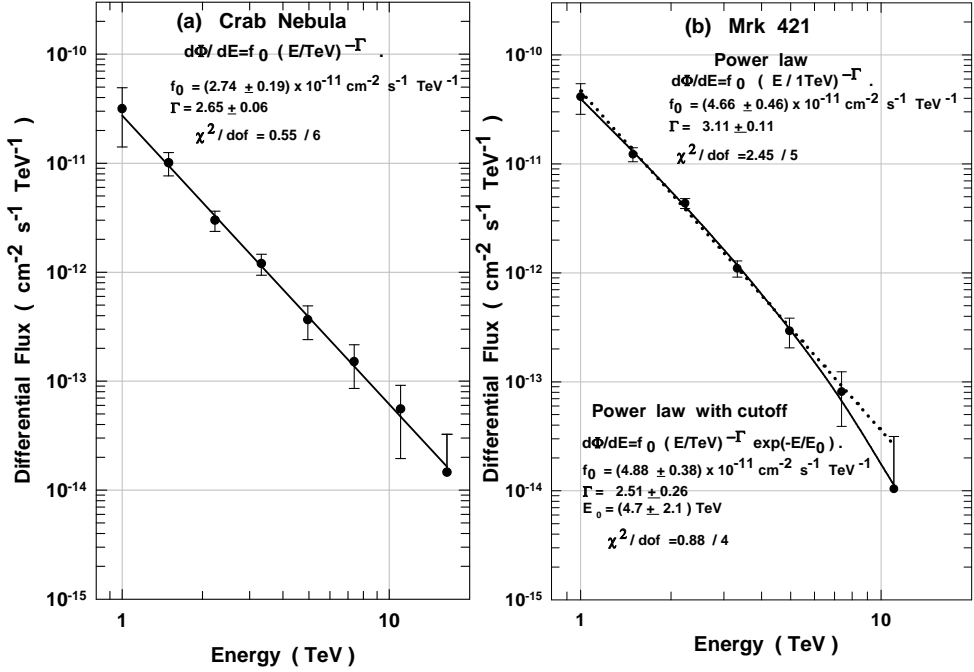}
\caption{ The differential  energy spectrum of the Crab Nebula as measured by the TACTIC  telescope. 
(b) Differential  energy spectrum of  Mrk 421 for the data collected between Dec. 27, 2005 - Feb. 07, 2006 when the source  was in a high state.}
\end{figure} 
\par
Turning to Mrk 421, since we know now that the source was in a high state during spell 2 and spell 3 observations, we have used this data alone to determine the time averaged  energy spectrum of Mrk 421. Fig.4b shows the  differential energy  spectrum  after  applying the Dynamic Supercuts to the combined  data of spell 2 and spell 3. A power law fit to the data ($d\Phi/dE=f_0 E^{-\Gamma}$) in the energy range 1-11 TeV yields $f_0=(4.66\pm0.46)\times 10^{-11}cm^{-2}s^{-1}TeV^{-1}$  and  $\Gamma=3.11\pm0.11$ with a $\chi^2/dof= 2.45/5$ (probability = 0.78). A power law with an exponential cutoff ($d\Phi/dE=f_0E^{-\Gamma}exp(-E/E_0$) was also tried and the results yield the following parameters of the  fit $f_0=(4.88\pm0.38)\times10^{-11}cm^{-2}s^{-1}TeV^{-1}$, $\Gamma=2.51\pm0.26$ and  $E_0=(4.7\pm2.1)TeV$ with a $\chi^2/dof= 0.88/4$ (probability = 0.93). The errors in the flux constant, the spectral index and cutoff energy are again  standard errors. While work on  understanding the telescope systematics is still in progress,  our preliminary   estimates for the Crab Nebula  spectrum indicate   that   the   systematic errors   in  flux   and  the  spectral index  are   
$<$ $\pm$ 40 $\%$ and $<$ $\pm$ 0.42, respectively.
\section{Discussion and Conclusions}
The observations of Mrk 421  carried out with the TACTIC imaging telescope at $\gamma$-ray energies of $\geq$ 1.2 TeV clearly indicate that the source was in a high state during Dec. 27, 2005 to Jan. 09, 2006  and Jan. 23, 2006 to Feb. 07, 2006  at a combined  average flux of $\sim$(1.04$\pm$0.14)CU. It is worth mentioning here that the preliminary results of the Whipple group [39] also indicate  that the source was in a high state during the period of their  observations  from Dec. 27, 2005 - Jan. 07, 2006 and Jan. 23, 2006 - Feb. 03, 2006. Furthermore,  our results  for spell 5 ( Mar. 19, 2006 to Mar. 30, 2006) are also consistent with the Whipple observations  from Mar. 23, 2006 to Apr. 04, 2006  when the source was  seen to flare at a  lower flux level. These observations clearly   indicate  that, despite  difference in the observation  time  between the two telescope systems,  Mrk 421  was  observed to be in  a high state by both the systems for  a prolonged duration,  somewhat  similar  to the flaring  episodes during Jan.-Feb., 2001  and Apr.- May, 2004.
\par
The energy spectrum of Mrk 421 as measured by the TACTIC imaging telescope in the energy  region  1-11 TeV is compatible with both a pure power law fit with an exponent of $\Gamma=3.11\pm0.11$
($\chi^2/dof= 2.45/5$ ; probability = 0.78) and a power law   with an exponential cut off  with $\Gamma=2.51\pm0.26$ and  $E_0=(4.7\pm2.1)TeV$ ($\chi^2/dof=0.88/4$; probability=0.93). However,  a systematic deviation  from a pure power law,  although not statistically very significant, is evident in Fig.4b  at the energy values of   7.4 TeV and 11.0 TeV.  Difficulties like limited $\gamma$-ray event statistics  coupled with rather large  error bars do not  allow us  to claim the cutoff feature at a high confidence level. Keeping in view  that the cutoff energy $E_0=(4.7\pm2.1)TeV$  inferred from  our  observations  is fairly consistent with the cutoff values of  $3.6(+0.4-0.3)_{stat}(+0.9-0.8)_{sys} TeV $ and $(4.3\pm0.3)TeV$ reported by  the HEGRA [11] and the VERITAS [16]
groups, respectively, it seems necessary to improve upon our analysis procedure so that the claim can be put on a firmer  basis. Furthermore, the results of the HESS group [7], based on 9 nights in Apr. and May 2004, also indicate that the time averaged energy spectrum of Mrk 421  is well described by a power law with index $\Gamma=2.1\pm0.1_{stat}\pm0.3_{sys}$ with  an exponential cutoff at  $3.1(+0.5-0.4)_{stat}\pm0.9_{sys}TeV$. The HESS results also indicate that the cutoff signature in the energy  spectrum  is intrinsic  to the source. 
\par    
In conclusion, we believe  that, despite the  limited sensitivity of the TACTIC telescope, the TeV light curve  presented here,  for nearly 5 months of extensive observations,  should  provide a useful  input   for the $\gamma$-ray astronomy community. Furthermore, we also  believe  that there is  considerable scope  for a TACTIC  like imaging telescope for monitoring AGNs on a long term basis. 
\section {Acknowledgements} 
The authors would like to convey their gratitude to all the concerned colleagues of the  Astrophysical Sciences  Division  for their contributions towards the instrumentation and observation aspects of the TACTIC telescope. 
We would also like to thank the anonymous referees for making several helpful suggestions.

\end{document}